\documentclass{elsarticle}
\usepackage[english]{babel}
\usepackage{hyperref}
\usepackage{amsmath}
\usepackage{caption}
\usepackage{subcaption}
\usepackage{graphicx}
\usepackage{natbib}
\usepackage[nameinlink]{cleveref}
\usepackage{siunitx}
\usepackage{isotope}
\usepackage{upgreek}
\usepackage{float}
\usepackage{orcidlink}
\usepackage{lineno}

\date{November 14, 2023}

\begin{document}
\title{A neutron trigger detector for pulsed reactor neutron sources}
\author[1]{Julian Auler\corref{cor1}\,\orcidlink{0009-0004-0554-0855}}
\author[2]{Dieter Ries\corref{cor2}\,\orcidlink{0000-0003-1663-6989}}
\author[3]{Bernd Ulmann\,\orcidlink{0000-0001-6967-8176}}
\author[1]{Evan Adamek}
\author[4]{Martin Engler\,\orcidlink{0000-0002-6110-8264}}
\author[1]{Martin Fertl\,\orcidlink{0000-0002-1925-2553}}
\author[4]{Konrad Franz\,\orcidlink{0009-0008-1822-3751}}
\author[1]{Werner Heil}
\author[4]{Simon Kaufmann\,\orcidlink{0009-0006-3149-6434}}
\author[1]{Niklas Pfeifer\,\orcidlink{0009-0003-0407-4023}}
\author[4]{Kim Ro\ss}
\author[4]{Alexandra Tsvetkov}
\author[4]{Noah Yazdandoost\,\orcidlink{0000-0001-6768-795X}}

\cortext[cor1]{Corresponding author: \texttt{juauler@uni-mainz.de}}
\cortext[cor2]{Corresponding author: \texttt{dieter.ries@psi.ch}}

\affiliation[1]{organization={Institute of Physics, Johannes Gutenberg University Mainz},
\postcode={55099},
\city={Mainz},
\country={Germany}}

\affiliation[2]{organization={Laboratory for Particle Physics, Paul Scherrer Institute (PSI)},
\postcode={5232},
\city={Villigen},
\country={Switzerland}}

\affiliation[3]{organization={Hochschule für Oekonomie und Management},
\postcode={60486},
\city={Frankfurt a.\,M.},
\country={Germany}}

\affiliation[4]{organization={Department of Chemistry - TRIGA site, Johannes Gutenberg University Mainz},
	\postcode={55099},
	\city={Mainz},
	\country={Germany}}

\begin{abstract}
 A variety of experiments investigating properties of neutrons can be performed at pulsed reactor neutron sources like the research reactor TRIGA Mainz.
 A typical problem faced by these experiments is the non-availability of a reliable facility-provided trigger signal in coincidence with the neutron production.
 Here we present the design and implementation of a neutron pulse detector that provides a coincident trigger signal for experimental timing with a relative precision of $\SI{4.5}{\milli\second}$.
\end{abstract}


 \maketitle 

\section{Introduction}
 Nuclear reactors of the TRIGA (Training, Research, Isotopes, General Atomics) type, like the research reactor TRIGA Mainz located at Johannes 
 Gutenberg University, can be operated in a pulsed mode:
 With the reactor initially operated at low steady 
 power, it is turned into a prompt super-critical state by shooting the so-called \emph{pulse rod} out of the core using pressurized air.
 The nuclear chain reaction starts an exponential power rise to a maximum of \SI{250}{\mega\watt_{th}} within a few
 milliseconds.
 The large prompt negative temperature coefficient of the TRIGA reactor's fuel material (ZrH) terminates the power excursion and thus forms a neutron pulse.
 Reactor pulses at the maximum power have a width of approximately \SI{30}{\milli\second}~(FWHM)~\cite{Frei2007}.

 A common task in controlling time-critical experiments based on the pulsed operation mode of the TRIGA reactor is the generation of a reliable trigger signal for every single pulse. 
 Using the reactor control systems for this purpose is problematic because of timing jitters in 
 the reactor control electronics and due to the pneumatic operation of the pulse rod, which is not pressure- or temperature-stabilized.
 These factors lead to uncertainties regarding the start time of the experiment and lead to a reduction in pulse-to-pulse reproducibility.
 
 At the research reactor TRIGA Mainz this applies in particular to the $\uptau$SPECT experiment~\cite{RoßPHD,KahlenbergPHD,KarchPHD}, which aims to measure the free neutron
 lifetime.
 To convert the thermalized neutrons from a reactor pulse to storable ultracold neutrons (UCNs) a solid deuterium based UCN source is used~\cite{Kahlenberg2017}.
 In $\uptau$SPECT UCNs are confined in a magnetic bottle.
 Plotting their number against the corresponding storage times provides the exponentially decaying storage curve, from which the neutron lifetime can be determined.
 Prior to each reactor pulse and experiment repetition with the reactor in a \emph{pulse ready} mode, a \emph{pulse ready signal} is sent from the reactor's control system to the control electronics of the experiment.
 This signal starts the timing sequence of the individual subsystems of the $\uptau$SPECT experiment with defined delay times being taken into account.
 The experiment then returns a \emph{pulse request signal} to the control room of the reactor which directly triggers the actual neutron pulse.

 To establish a facility-independent trigger signal at the research reactor TRIGA Mainz a new trigger signal generation method has been developed and implemented.
 Here, facility-independent means that the generation of the trigger signal should be 1) independent of the mechanical and electronic control systems 2) independent of the pneumatic pulse rod system and 3) independent of reactor-physical effects such as reactivity, steady state power or reactor poisoning.
 A dedicated neutron detector, called \emph{Trigger Detector} in the following, registers the thermal neutrons of the pulse, which arrive basically instantaneously at the experiment compared to the storable UCNs with very low kinetic energy, and generates a digital trigger signal, which is used for the timing of the experiment only.
 Such a trigger signal is physically correlated to neutron production and completely independent of systematic effects from the operation of the facility. 

 In addition to its intended use at the research reactor TRIGA Mainz or comparable neutron sources, the Trigger Detector could possibly also be used at accelerator-driven pulsed neutron sources, such as the Spallation Neutron Source (SNS) or the European Spallation Source (ESS). 
 This type of pulsed neutron source has a much higher pulse repetition rate, which for SNS and ESS, e.g., is \SI{60}{\hertz} and \SI{14}{\hertz} respectively~\cite{SNS,ESS}.

 \section{Hardware}
 
 \subsection{Overview of trigger signal generation}
\begin{figure}
   \centering
   \vspace*{-5mm}
   \includegraphics[width=1.0\textwidth]{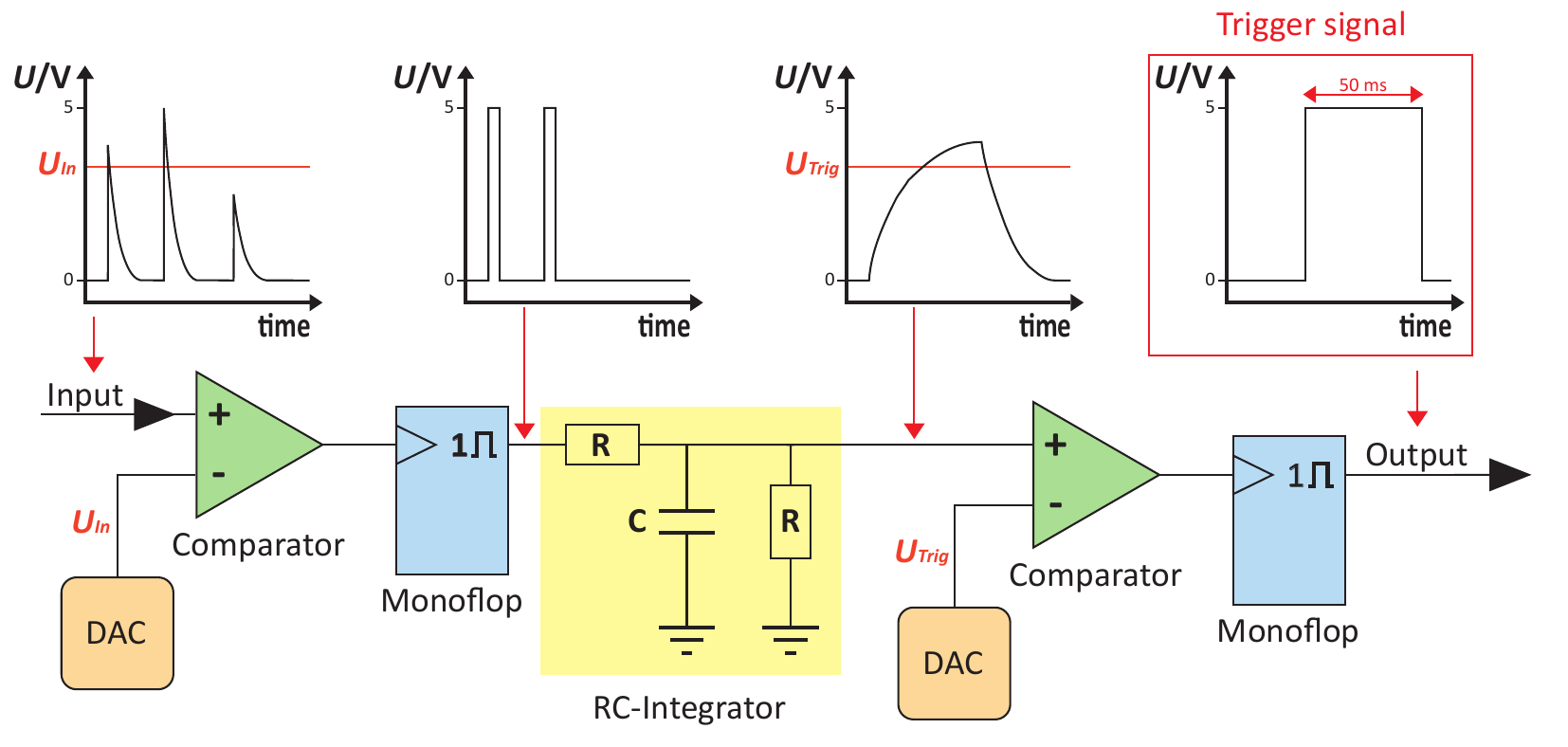}
   \vspace*{-5mm}
   \caption{Schematic overview of trigger signal generation with the presented neutron pulse trigger detector. Here $U_{In}$ refers to the input threshold of the first comparator and $U_{Trig}$ to the trigger threshold of the second comparator.}
   \label{high_level_scheme}
  \end{figure}

 The principle of trigger signal generation for the Trigger Detector is shown in \cref{high_level_scheme}.
 Individual neutron events detected by a neutron sensitive multilayer detector (see \cref{Neutron_sensitive_multilayer_detector}) serve as the input to the trigger electronics.
 In a first step these events with rather arbitrary width and amplitude are converted into uniform square pulses, which are generated by a monoflop after a first comparator has discriminated the individual events by their amplitude.
 A downstream RC integrator circuit is then charged for each pulse generated by the monoflop.
 Once the charge on the RC combination exceeds the threshold of a second comparator, another monoflop is triggered, producing a \SI{50}{\milli\second} long trigger signal as the primary output of the trigger detector.
 The corresponding input and trigger threshold for the first and second comparator, respectively, are generated by two digital to analog converters (DACs) and can be adjusted.
 The trigger electronics are described in detail in \cref{Neutron_trigger_detector_circuit} and the setting as well as the importance of the input and trigger threshold in \cref{The_Trigger_Detector}.
 
 \subsection{Neutron sensitive multilayer detector}\label{Neutron_sensitive_multilayer_detector}
 
 \begin{figure}
   \centering
   \vspace*{-5mm}
   \includegraphics[width=.5\textwidth]{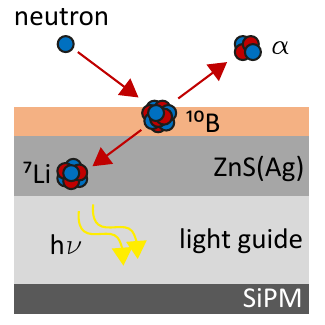}
   \vspace*{-5mm}
   \caption{Principle of operation of a multilayer detector for neutrons, consisting of a \isotope[10]{B}
   top layer, a scintillation layer of ZnS(Ag) and a light guide as supporting layer, followed by a
   silicon photomultiplier as photosensor. Although this detector design is typically used to detect UCNs, it is used in the Trigger Detector to detect much faster, thermal neutrons.}
   \label{pic_multilayer_sensor}
  \end{figure}
  
 This detector, intended to detect thermal neutrons, is based on a multilayer design (see \cref{pic_multilayer_sensor}) typically used for UCNs, and employs the $\isotope[10]{B}(\text{n},\upalpha)\isotope[7]{Li}$ reaction, as e.g. also reported in ~\cite{WANG201530}.
 Compared to other techniques for the detection of neutrons\footnote{\cite{Pie20} gives a nice overview of techniques for the detection of neutrons in different energy ranges and their applications.} the choice of $\isotope[10]{B}$ in a multilayer design allows a very simple and cost-effective setup of the Trigger Detector.
 In addition, there is a synergy with UCNs through the connection to the $\uptau$SPECT experiment, which strongly influences the choice of this technique.
  
 When a neutron is captured within the $\isotope[10]{B}$ top layer an $\upalpha$-particle and a $\isotope[7]{Li}$-nucleus are emitted in opposite directions.
 The particle emitted towards the scintillating ZnS(Ag) layer causes light to be emitted, which is then transmitted to a photosensor. 
 The optimal $\isotope[10]{B}$-layer thickness is determined by two counteracting demands.
 The layer has to be thick enough to absorb the neutron with high probability but thin enough to allow the charged ejectiles ($\upalpha$ or $\isotope[7]{Li}$) to reach the ZnS(Ag) layer.
 While for UCNs a thickness of the $\isotope[10]{B}$ top layer of about \SI{100}{\nano\meter} is sufficient~\cite{WANG201530}, a higher neutron absorption efficiency is to be expected for thermal neutrons with larger $\isotope[10]{B}$ layer thicknesses.
 The maximum ion range of the charged products from the $\isotope[10]{B}(\text{n},\upalpha)\isotope[7]{Li}$ reaction in $\isotope[10]{B}$ solid films and in ZnS are reported in~\cite{WANG201530} and is in the order of a few micrometers, which is an upper limit for the possible thickness of the $\isotope[10]{B}$ top layer.
 At the same time, this means that a thickness of a few micrometers is sufficient for the scintillating ZnS(Ag) layer~\cite{WANG201530}.

 For the Trigger Detector a scintillator foil\footnote{EJ-440 - \textit{ELJEN TECHNOLOGY}}, consisting of a \SI{0.25}{\milli\meter} thick polyester supporting layer (light guide) and a one-sided coating of ZnS(Ag) as scintillation layer, is used.
 The scintillation light has a wavelength distribution centered at \SI{450}{\nano\meter} and a decay time constant of \SI{200}{\nano\second}~\cite{eljen}.
 The scintillator foil is coated with a $\isotope[10]{B}$ top layer of about \SI{80}{\nano\meter} thickness\footnote{\textit{CDT CASCADE Detector Technologies GmbH}}.
 By choosing this $\isotope[10]{B}$ top layer thickness, which again is strongly influenced by the synergy with UCN experiments, and by not adapting a larger layer thickness more suitable for thermal neutrons, the detection efficiency of the Trigger Detector will be decreased.
 Since the aim of the Trigger Detector is the detection of the reactor pulse as a whole and not the counting of individual neutrons, the detection efficiency is not a critical parameter and the described decrease in efficiency is acceptable.

 A silicon photomultiplier\footnote{C-Series 6MM, MICROFC-60035-SMT-TR1 - \textit{Semiconductor Components Industries}} (SiPM) is used as photosensor.
 The SiPM consists of a grid of independent Single Photon Avalanche Diodes, each with its own quench resistor referred to as microcell.
 The SiPM used here includes 18980 microcells covering an active area of \SI{36}{\milli\meter\squared}.
 It has a spectral sensitivity of \SI{300}{\nano\meter} to \SI{950}{\nano\meter} with a peak sensitivity at \SI{420}{\nano\meter}. 
 The operating voltage of the SiPM can be determined from the breakdown voltage of the diode and the permitted overvoltage.
 Here the bias voltage is \SI{-27}{\volt}, with the output current of the power supply limited to \SI{15}{\milli\ampere} to protect the sensor~\cite{microc}. 
 In the Trigger Detector, the SiPM is biased and read out using the cathode (C) and anode (A) contacts, not using the capacitively coupled fast output (F)~\cite{and9770}.
 The scintillator foil is attached to the SiPM using transparent silicone paste\footnote{KORASILON high viscosity - \textit{Kurt Obermeier GmbH \& Co. KG}}.
 \subsection{Neutron trigger detector circuit}\label{Neutron_trigger_detector_circuit}
\begin{figure}[h]
  \centering
  \includegraphics[width=.75\textwidth]{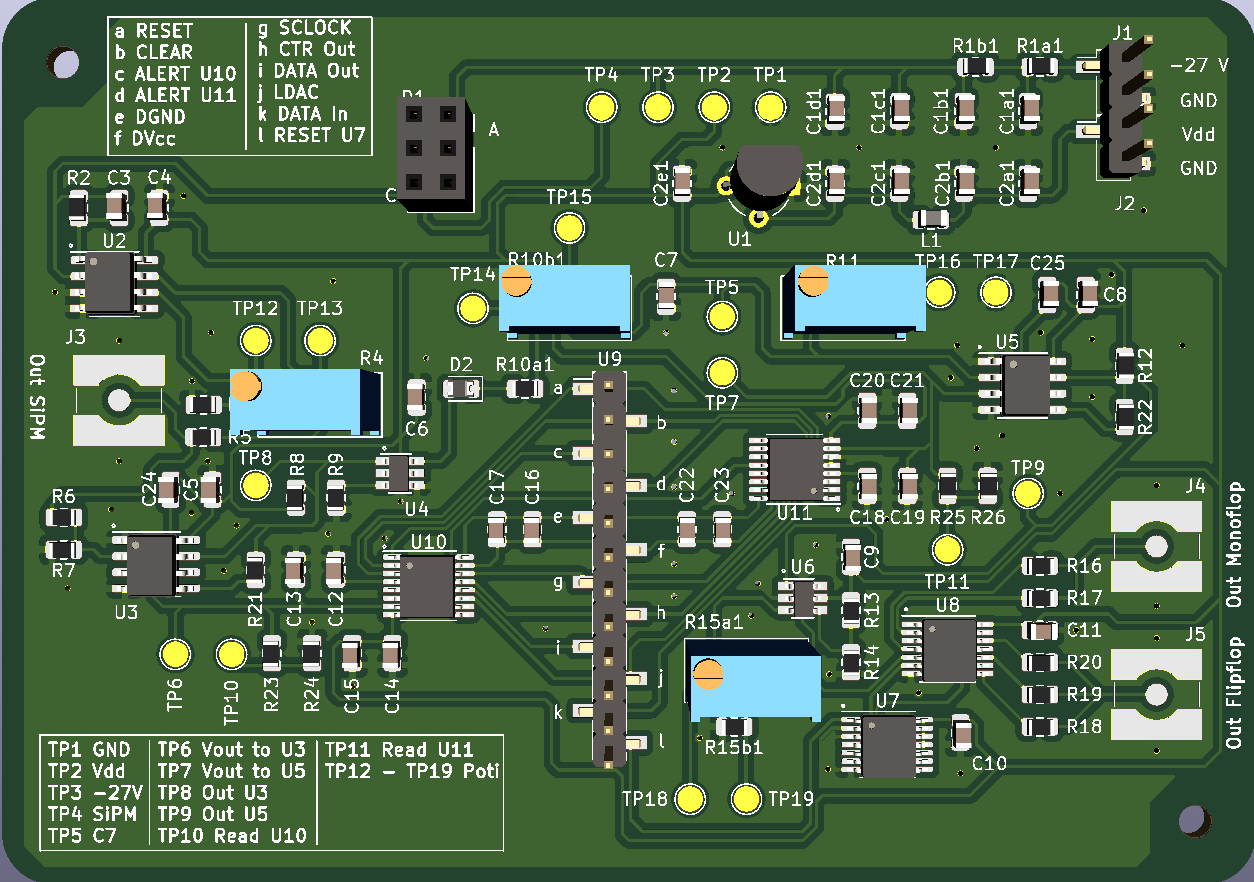}
  \caption{Rendering of the Trigger Detector board.}
  \label{pic_rendered_board}
 \end{figure}

\begin{figure}
  \centering
  \includegraphics[width=0.8\paperwidth , angle=270]{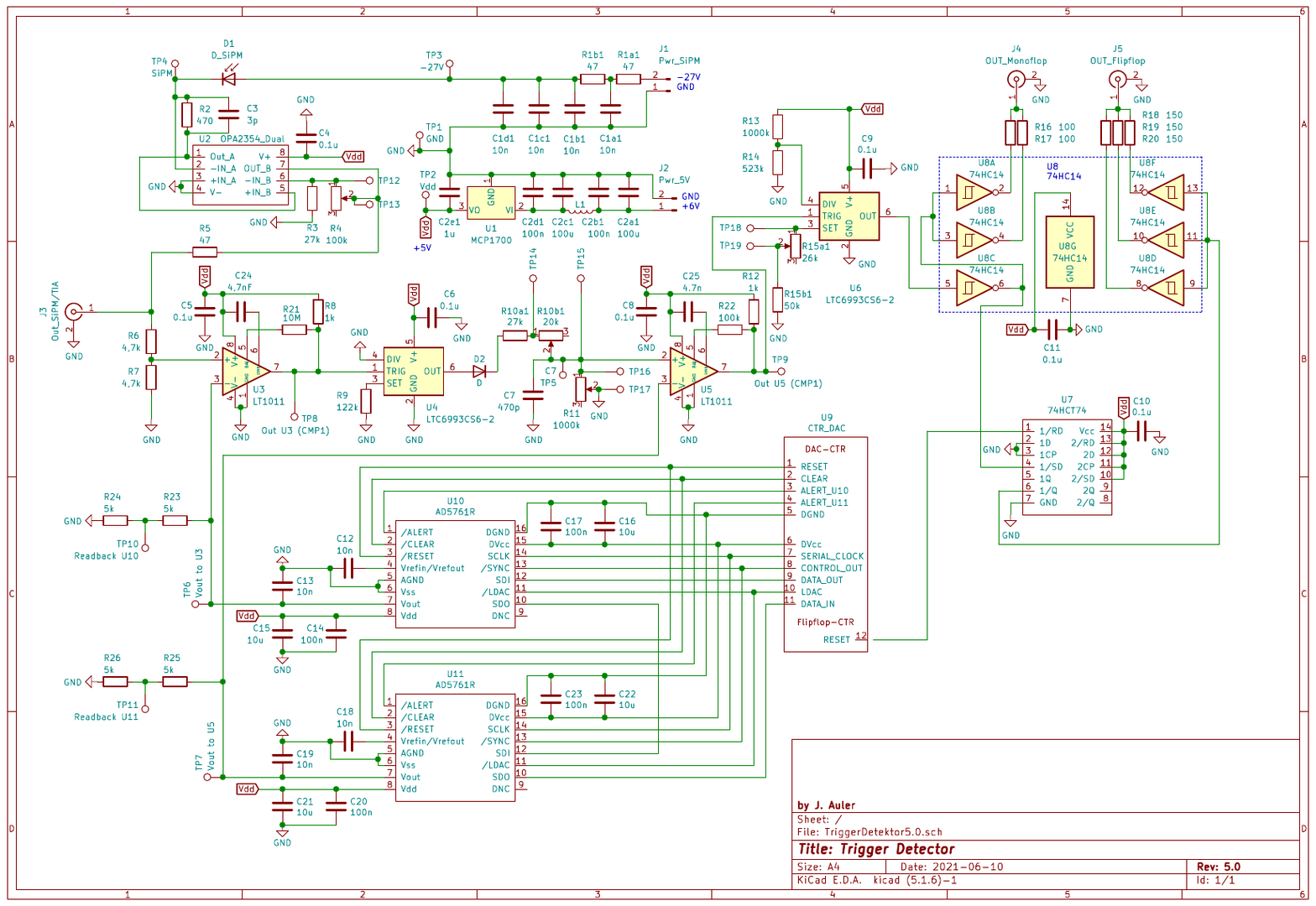}
  \caption{Schematic of the neutron trigger detector circuit.}
  \label{pic_schematic}
 \end{figure}

 A rendering of the neutron trigger detector electronics circuit as well as the schematic is shown in \cref{pic_rendered_board} and \cref{pic_schematic}.
 The Trigger Detector printed circuit board (PCB) requires two operating voltages, \SI{-27}{\volt} for the SiPM bias voltage and a \SI{6}{\volt} supply voltage for the detector electronics.
 The SiPM is equipped with an upstream filter stage consisting of two low-pass RC filters (R1a1/C1a1 and R1b1/C1b1) and two decoupling capacitors (C1c1 and C1d1) according to the manufacturers suggestions~\cite{and9782}.
 The \SI{6}{\volt} supply voltage is fed to a linear voltage regulator\footnote{MCP1700-5002E/TO - \textit{Microchip Technology Inc.}} U1 yielding \SI{5}{\volt} at its output.
 All active components have a local \SI{0.1}{\micro\farad} decoupling capacitor.


 The current induced by the photon interaction in the SiPM is converted into a corresponding voltage by means of a transimpedance amplifier (TIA) consisting of the first half of the dual operational amplifier\footnote{OPA2354A1DDA - \textit{Texas Instruments Incorporated}} U2.
 The gain is determined mainly by R2.
 The parallel capacitor C3 introduces a tiny amount of integrating behaviour thus smoothing the resulting output signal.
 The TIA is followed by a non-inverting amplifier, the second half of U2.
 The adjustable gain of this stage is set by the voltage divider consisting of R3 and R4 which is implemented as a trimmer.

 The output signal of this amplifier consists of many individual spikes cor\-re\-spond\-ing to single events from the SiPM.
 It is fed to connector J3 (Out 3 - SiPM) for debugging purposes.
 It is also fed into a comparator\footnote{LT1011CS8 - \textit{Analog Devices Inc.}} U3 through a voltage divider consisting of R6 and R7. 
 The comparator converts all incoming pulses exceeding a certain threshold to pulses of equal height which are used to trigger a monoflop\footnote{LTC6993CS6-2 - \textit{Analog Devices Inc.}} U4.
 Thus a pulse from the SiPM with rather arbitrary width and amplitude is converted into a pulse of fixed height and width according to \cref{eq1}\,\cite{ltc6993}:
 \begin{equation}
	 t_{\text{OUT}} = \dfrac{N_{\text{DIV}} \cdot R_{\text{SET}}}{50~\text{k}\Omega}\cdot
	 \SI{1}{\micro\second}
     \label{eq1}
 \end{equation}
 
 $N_{\text{DIV}}$ is the divider ratio of the monoflop and set to $1$ by connecting pin 4 to ground.
 $R_{\text{SET}} = \text{R9} = \SI{122}{\kilo\ohm}$ then determines the pulse width as $t_{\text{OUT}}(U4) = \SI{2.44}{\micro\second}$.

 These pulses are then fed into a simple RC integrator circuit consisting of R10a1, R10b1, C7, and R11.
 The diode D2 makes sure that the RC combination is not discharged back into the monoflop while the trimmer R10b1 sets the integration constant, i.\,e. the amount by which the RC combination is charged for each pulse generated by the monoflop.
 The trimmer R11 controls the discharge of the RC combination effectively implementing a ``leaky'' integrator.
 The output of this integrator circuit is then fed to another comparator U5 which generates an output pulse whenever the voltage at the RC combination exceeds its threshold.
 This pulse triggers yet another monoflop U6 which generates the output trigger pulse.
 The monoflop is set to a pulse width of \SI{50}{\milli\second} by connecting its DIV input to the voltage divider R13/R14 and by the resistor combination R15b1 and R15a1 connected to the SET input.
 
 \begin{table}[h]
     \centering
     \begin{tabular}{|ll|l|}
          \hline
	  &&\\
         Board (U9)&                  & Teensy     \\
	 &&\\
          \hline
         a     & RESET            &  -         \\
         b     & CLEAR            &  -         \\
         c     & ALERT U10        &  -         \\
         d     & ALERT U11        &  -         \\
         e     & DGND             &  G Pin     \\
         f     & DVcc             &  \SI{3.3}{\volt} Pin \\
         g     & SCLOCK           &  Pin 27    \\
         h     & CTR Out          &  Pin 38    \\
         i     & DATA Out         &  Pin 26    \\
         j     & LDAC             &  -         \\
         k     & DATA In          &  Pin 39    \\
         l     & RESET U7         &  Pin 37    \\
         TP10  & READBACK U10     &  Pin 22    \\
         TP11  & READBACK U10     &  Pin 23    \\
         TP1   & GND              &  G Pin     \\
         TP2   & Vdd              &  \SI{5}{\volt} Pin   \\
         \hline
     \end{tabular}
     \caption{Connections from the Trigger Detector board to the Teensy 4.1 development board.}
     \label{tab:1}
 \end{table}
 
 This output signal is fed through the Schmitt trigger inverter\footnote{MM74HC14YTTR - \textit{STMicroelectronics}} U8C to the two inverters U8A/B and is available at jack J4 (Out 1 - monoflop).
 It also sets a flipflop\footnote{MM74HCT74MTC - \textit{Semiconductor Components Industries}} U7.
 The output of this flipflop is also buffered by three paralleled Schmitt trigger inverters U8D/E/F and fed to jack J5 (Out 2 - flipflop).
 The relevant thresholds of the two comparators U3 and U5 are generated by two daisy chained DACs\footnote{AD5761RARUZ - \textit{Analog Devices Inc.}} U10 and U11, which are controlled by an attached Teensy 4.1 microcontroller board\footnote{https://www.pjrc.com/store/teensy41.html}. 
 These threshold parameters can also be read out by the test jacks TP10 and TP11 which are connected to voltage dividers to ensure that the voltage does not exceed a maximum value of \SI{3.3}{\volt}.
 The microcontroller also resets the flipflop U7 by means of the RESET control line.
 The connections between the Trigger Detector board and the Teensy 4.1 are described in \cref{tab:1}.
 Communication with the microcontroller takes place via Ethernet.
 
 \subsection{The Trigger Detector}\label{The_Trigger_Detector}
 \begin{figure}[h]
 	\centering
 	\begin{subfigure}{0.35\textwidth}
 		\includegraphics[width=\textwidth]{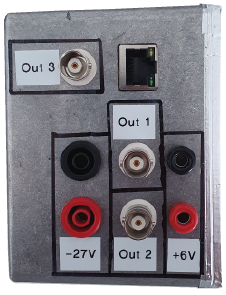}
 		\caption{}
 		\label{DetectorBoxA}
 	\end{subfigure}
 	\hfill
 	\begin{subfigure}{0.63\textwidth}
 		\includegraphics[width=\textwidth]{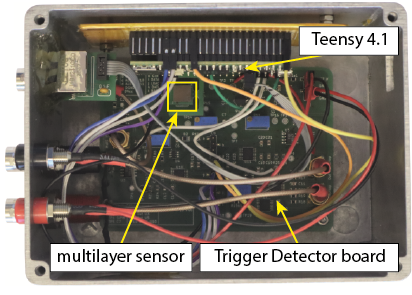}
 		\caption{}
 		\label{DetectorBoxB}
 	\end{subfigure} 	
 	\caption{Aluminum enclosure of the Trigger Detector: (a) connectors for both supply voltages, the three output signals Out1 (monoflop), Out2 (flipflop) and Out3 (SiPM) and the Ethernet connection; (b) inside view of the Trigger Detector.}
 	\label{housing_picture}
 \end{figure}
 
 The Trigger Detector board and the Teensy 4.1 development board are mounted inside an aluminium enclosure (\cref{housing_picture}) which has all inputs and outputs of the detector on the left hand side.
 During operation, the detector is wrapped up in a black double layer textile bag to exclude ambient light. 
 Using a HTTP based graphical user interface running on a webserver on the Teensy 4.1, the two thresholds \cref{eq2} and \cref{eq3} for comparators U3 and U5 can be easily set by means of the two DACs U10 and U11, where the setting parameters $N_{\text{SET}}(U10)$ and $N_{\text{SET}}(U11)$ are integers in the interval $[0,4095]$.
 Linearity measurements of each DAC were performed to guarantee that the threshold voltage corresponds directly to the parameter $N_{\text{SET}}$ (\cref{eq2} and \cref{eq3}).
 
  \begin{equation}
     U_{\text{SET}}(U10) = \SI{1.19}{\milli\volt} \cdot N_{\text{SET}}(U10) + \SI{0.50}{\milli\volt}
     \label{eq2}
 \end{equation}
  \begin{equation}
     U_{\text{SET}}(U11) = \SI{1.19}{\milli\volt} \cdot N_{\text{SET}}(U11) + \SI{0.26}{\milli\volt}
     \label{eq3}
 \end{equation}
 
 These two threshold voltages can be independently read back as $N_\text{READ}(U10)$ and $N_\text{READ}(U11)$ according to \cref{eq4} and \cref{eq5} (the aforementioned voltage dividers have to be taken into account here):
 
 \begin{equation}
     U_{\text{READ}}(U10) = \SI{0.78}{\milli\volt} \cdot N_\text{READ}(U10) + \SI{1.49}{\milli\volt}
     \label{eq4}
 \end{equation}
  \begin{equation}
     U_{\text{READ}}(U11) = \SI{0.78}{\milli\volt} \cdot N_\text{READ}(U11) + \SI{1.54}{\milli\volt}
     \label{eq5}
 \end{equation}

\begin{figure}[h]
 	\centering
 	\begin{subfigure}{0.49\textwidth}
 		\includegraphics[width=\textwidth]{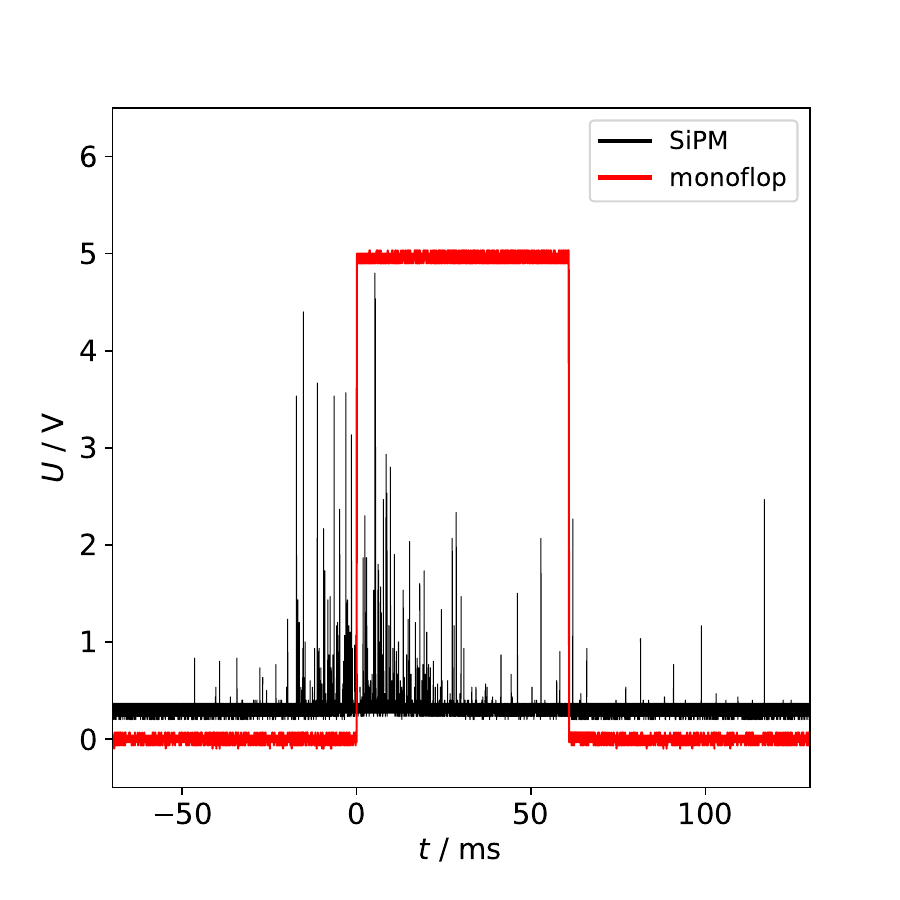}
 		\caption{}
 		\label{monoflop_pictureA}
 	\end{subfigure}
 	\hfill
 	\begin{subfigure}{0.49\textwidth}
 		\includegraphics[width=\textwidth]{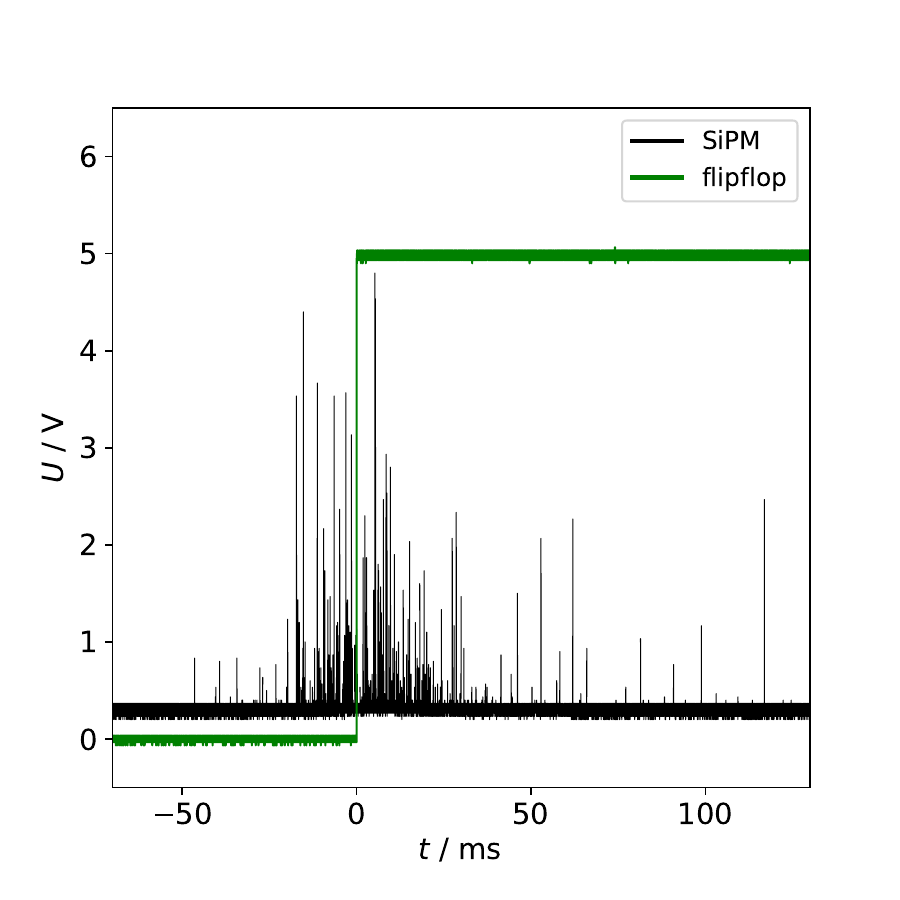}
 		\caption{}
 		\label{monoflop_pictureB}
 	\end{subfigure} 	
 	\caption{Examples of trigger signals generated during a reactor pulse with the trigger detector set to $N_{\text{SET}}(U10) = 200$ and $N_{\text{SET}}(U11) = 1500$. Initial signal at the SiPM output (black) and subsequently generated trigger signal: (a) at the monoflop output (red); (b) at the flipflop output (green). Time scale relative to the rising edge of the trigger signal.}
 	\label{monoflop_picture}
 \end{figure}

 The threshold parameters are critical to get good trigger signals based on a reactor pulse.
 First, an input threshold must be chosen which is high enough to discriminate background from actual neutron counts.
 A value of $N_{\text{SET}}(U10) = 200$ was empirically determined for our setup and reactor by increasing the threshold until the background no longer produced trigger signals.
 In the same way, the trigger threshold has to be chosen carefully to avoid spurious events generating a trigger signal while simultaneously guaranteeing that the trigger signal is reliably generated by a real neutron pulse.
 An example of the trigger signal is given for the monoflop and the flipflop output (\cref{monoflop_picture}) generated by a single reactor pulse.
 Unlike the monoflop, the flipflop remains in its activated state after triggering and must be actively reset via software.
 In this way, signals of any length can be generated in parallel to the monoflop trigger signal, e.g. as a defined veto signal that prevents the acceptance of another trigger and can therefore be used for experiment control.
 
 \section{Results}
 \subsection{Reactor pulse simulation}
 The circuit was simulated using LTspice\footnote{LTspice XVII(x64)(17.0.33.0) - \textit{Analog Devices, Inc.}}.
 The input pulses in the simulation are based on the typical signal shape of individual neutron events\footnote{The values used are assumptions made on the basis of an oscilloscope measurement.} observed on the so-called \emph{thermal column} of the TRIGA reactor\footnote{The thermal column of the TRIGA reactor is a beam port that has a high thermal neutron flux due to a graphite moderator.}.
 A single neutron event is assumed to have a rise time of \SI{1}{\micro\second} and a fall time of \SI{20}{\micro\second} with an amplitude of \SI{5}{\volt}.
 3000 such individual events are fed into the simulation corresponding to a reactor pulse of \SI{30}{\milli\second} duration.
 The input and trigger thresholds are $U_{\text{SET}}(U10) = \SI{240}{\milli\volt}$ and $U_{\text{SET}}(U11) = \SI{1785}{\milli\volt}$ corresponding to the threshold parameters $N_{\text{SET}}(U10) = 200$ and $N_{\text{SET}}(U11) = 1500$.

\begin{figure}[h]
 	\centering
 	\begin{subfigure}{0.49\textwidth}
 		\includegraphics[width=\textwidth]{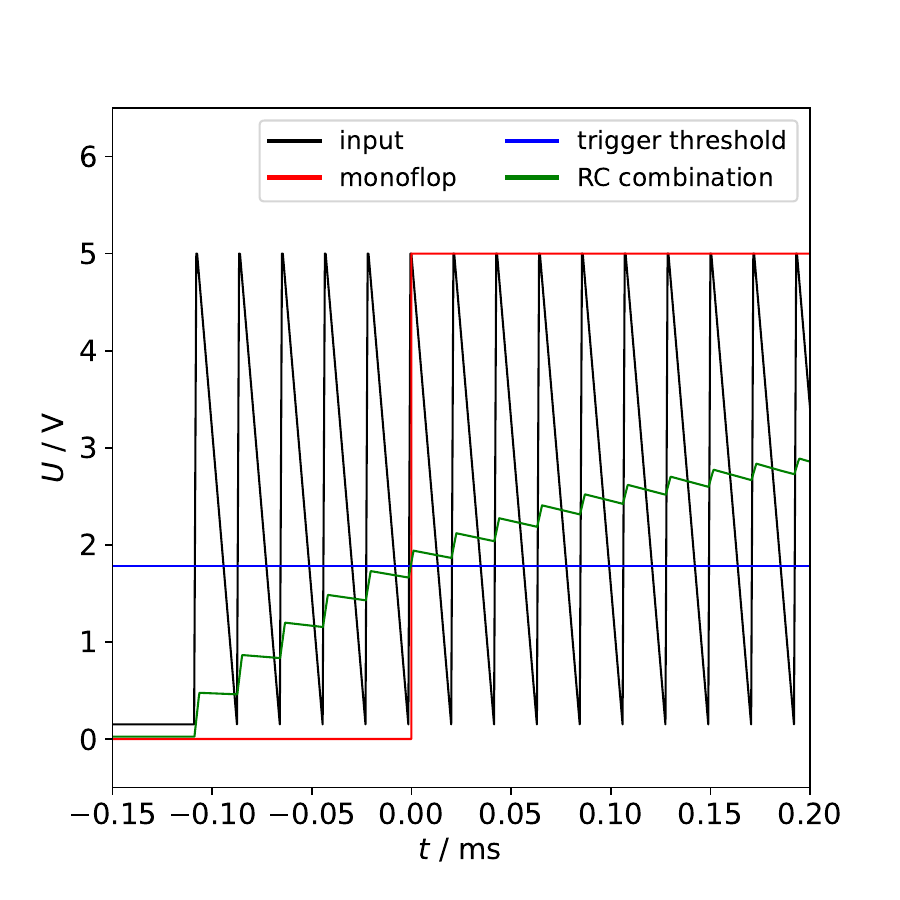}
 		\caption{}
 		\label{ledA}
 	\end{subfigure}
 	\hfill
 	\begin{subfigure}{0.49\textwidth}
 		\includegraphics[width=\textwidth]{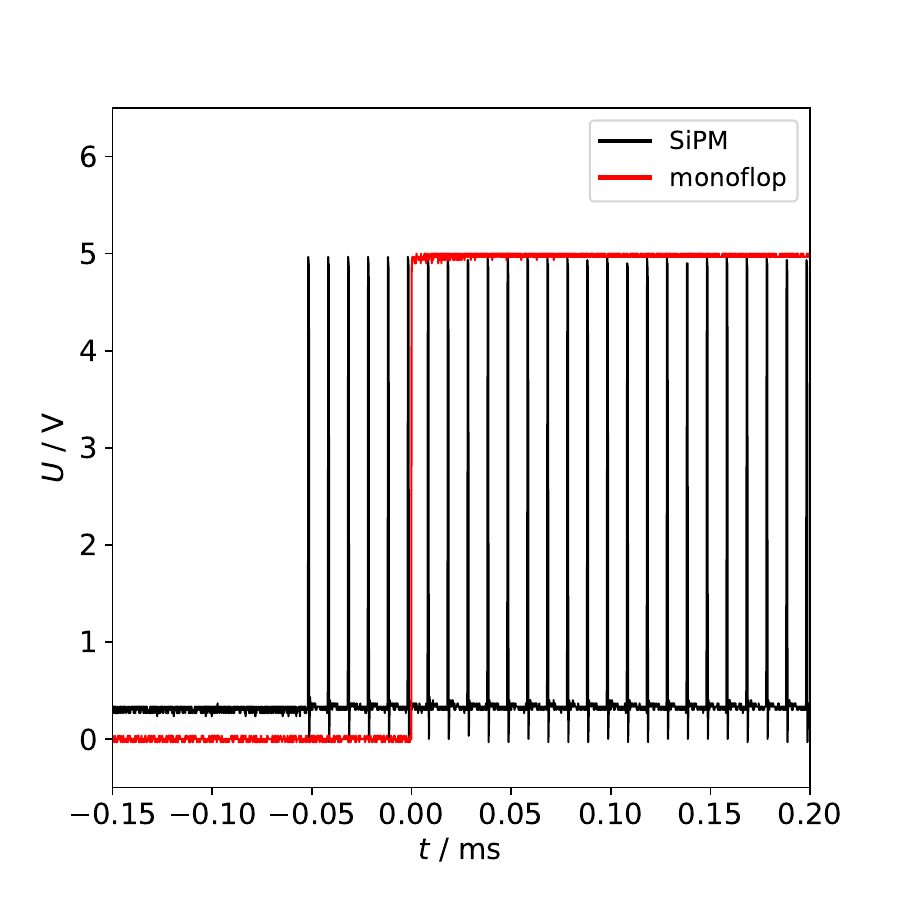}
 		\caption{}
 		\label{ledB}
 	\end{subfigure} 	
 	\caption{(a) Simulation run of the Trigger Detector circuit with the input signal (black) charging the RC combination (green) which then triggers the output signal (red) when the trigger threshold (blue) is exceeded (thresholds set to $U_{\text{SET}}(U10) = \SI{240}{\milli\volt}$ and $U_{\text{SET}}(U11) = \SI{1785}{\milli\volt}$). (b) Actual circuit with pulsed LED as signal source to simulate incoming neutrons. Threshold set identical to the simulation by the threshold parameters $N_{\text{SET}}(U10) = 200$ and $N_{\text{SET}}(U11) = 1500$. Time scale relative to the rising edge of the trigger signal.}
 	\label{led}
 \end{figure}

 This simulation is compared to the actual behavior of the circuit which is fed light pulses from a LED connected to a suitable signal generator.
 The comparison shows that the simulation and the LED measurement are in good agreement for the given setting (see \cref{led}).
 Furthermore, the simulation illustrates the operating principle of the circuit described above.
 
 \subsection{Optimal trigger threshold for different locations}\label{PositionDependency}
 The position of the Trigger Detector was changed relative to the TRIGA reactor in order to find the optimal placement and to study the effect of the changed neutron flux on the trigger threshold $N_{\text{SET}}(U11)$.
 For this purpose, reactor pulses were recorded at distances of approximately four, two, and one meters from beam port D of the reactor respectively\footnote{From the concrete biological shield of the TRIGA reactor at beam port D to the reactor core is another 302~cm \cite{Karch14}.}.
 In all three positions series of measurements were performed with the input threshold set to $N_{\text{SET}}(U10) = 200$, while the trigger threshold $N_{\text{SET}}(U11)$ was varied.
 The trigger signal was reliably triggered for trigger threshold values of up to $N_{\text{SET}}(U11) = 450$ (at \SI{4}{\meter}), $N_{\text{SET}}(U11) = 950$ (at \SI{2}{\meter}) and $N_{\text{SET}}(U11) = 1600$ (at \SI{1}{\meter}).
 Due to the higher neutron flux near the reactor, the number of individual neutron events that are detected and included in the triggering is higher than in more distant positions.
 As a result, the trigger threshold $N_{\text{SET}}(U11)$ can be set to high values at positions close to the reactor in order to detect the reactor pulse more reliably and to efficiently prevent spurious events from generating the trigger signal.
 Accordingly, the optimal parameters for the position closest to the reactor directly in front of beam port D (at \SI{0}{\meter}) were determined empirically as $N_{\text{SET}}(U10) = 200$ and $N_{\text{SET}}(U11) = 3000$. 
 
 \subsection{Comparison to Cherenkov radiation}
 The production of neutrons in a nuclear reactor coincides with the occurrence of Cherenkov radiation.
 For this reason, and because of the ease of observability due to the open swimming pool design of the research reactor TRIGA Mainz, Cherenkov radiation is well suited for independent comparison with the Trigger Detector\footnote{Cherenkov radiation is not used to generate a trigger signal at the research reactor TRIGA Mainz since it is not independent of reactor-physical effects. For other pulsed sources the occurrence and the ease of observability of Cherenkov radiation are not necessarily present.}.
 For this purpose, the output signal of the Trigger Detector was correlated with the time structure of the Cherenkov radiation emitted during a reactor pulse, which was detected with a photodiode from the top of the reactor pool.
 
 \begin{figure}[H]
 	\centering
 	\includegraphics[width=0.5\textwidth]{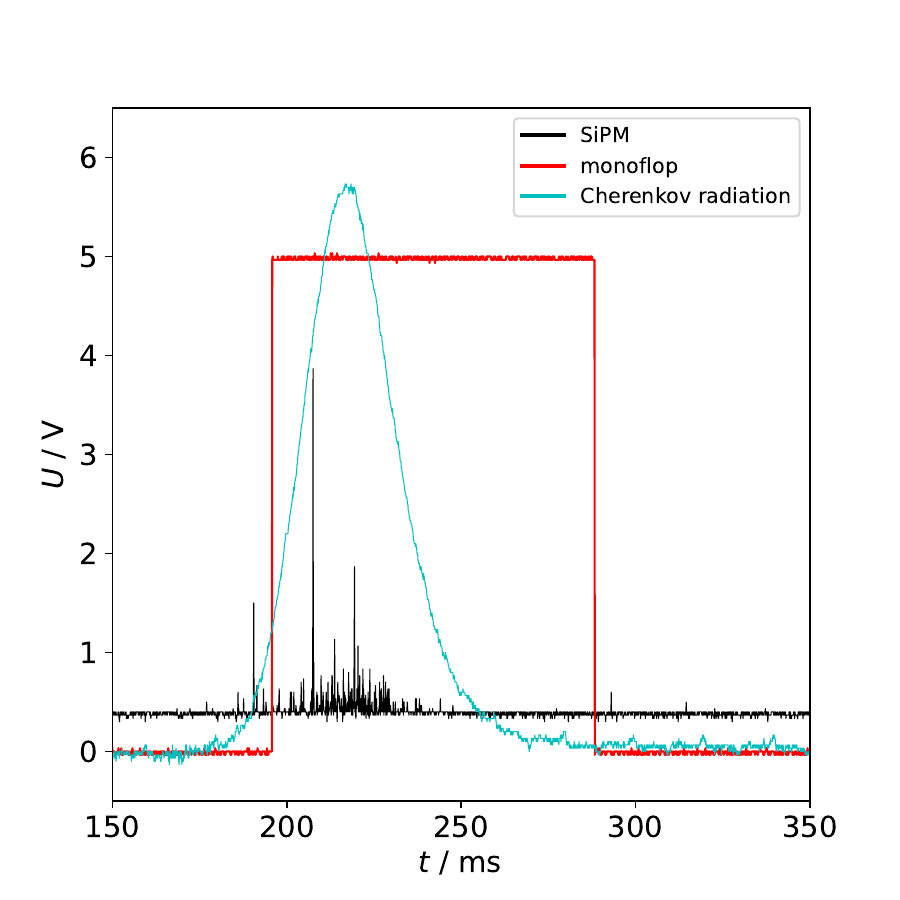}
 	\caption{Comparison of Cherenkov radiation (turquoise) and output of the Trigger Detector (red). The black pulses are individual events detected by the SiPM. Thresholds were set to $N_{\text{SET}}(U10) = 200$ and $N_{\text{SET}}(U11) = 3000$. Time scale relative to the TTL signal triggered in pulsed mode when the pulse rod hits the stop.}
 	\label{MonoCherenkov}
 \end{figure}  

 For this comparison the Trigger Detector was positioned directly in front of beam port D of the reactor using the optimal parameters $N_{\text{SET}}(U10) = 200$ and $N_{\text{SET}}(U11) = 3000$ for this position.
 Cherenkov radiation and output signal of the Trigger Detector are recorded simultaneously using an oscilloscope\footnote{SDS5104X - \textit{SIGLENT Technologies}} as shown in \cref{MonoCherenkov}.
 Here, the timescale is relative to a transistor-transistor logic (TTL) signal that is triggered in pulsed mode when the pulse rod hits its stop.
 The figure shows a deviation of the length of the trigger signal from the intended \SI{50}{\milli\second} (see \cref{Neutron_trigger_detector_circuit}).
 
 Moreover, the length of the monoflop output is not reproducible from pulse to pulse, but varies up to about $\SI{90}{\milli\second}$.
 This behavior was observed only during beam times at the reactor.
 Off-line studies in which a flashing LED is used, result in an average trigger signal length of $\overline{t} = \SI{48.87\pm0.03}{\milli\second}$.
 This suggests that the trigger electronics are affected by the neutron and gamma radiation to which the trigger detector at the reactor beam port is exposed during operation.
 Since the rising edge is most important for the timing of an experiment, the varying length of the trigger signal does not limit the function of the Trigger Detector.
 However, the behavior described above will be further investigated, especially with respect to radiation damage.
 
 \begin{figure}[h]
 	\centering
 	\includegraphics[width=0.75\textwidth]{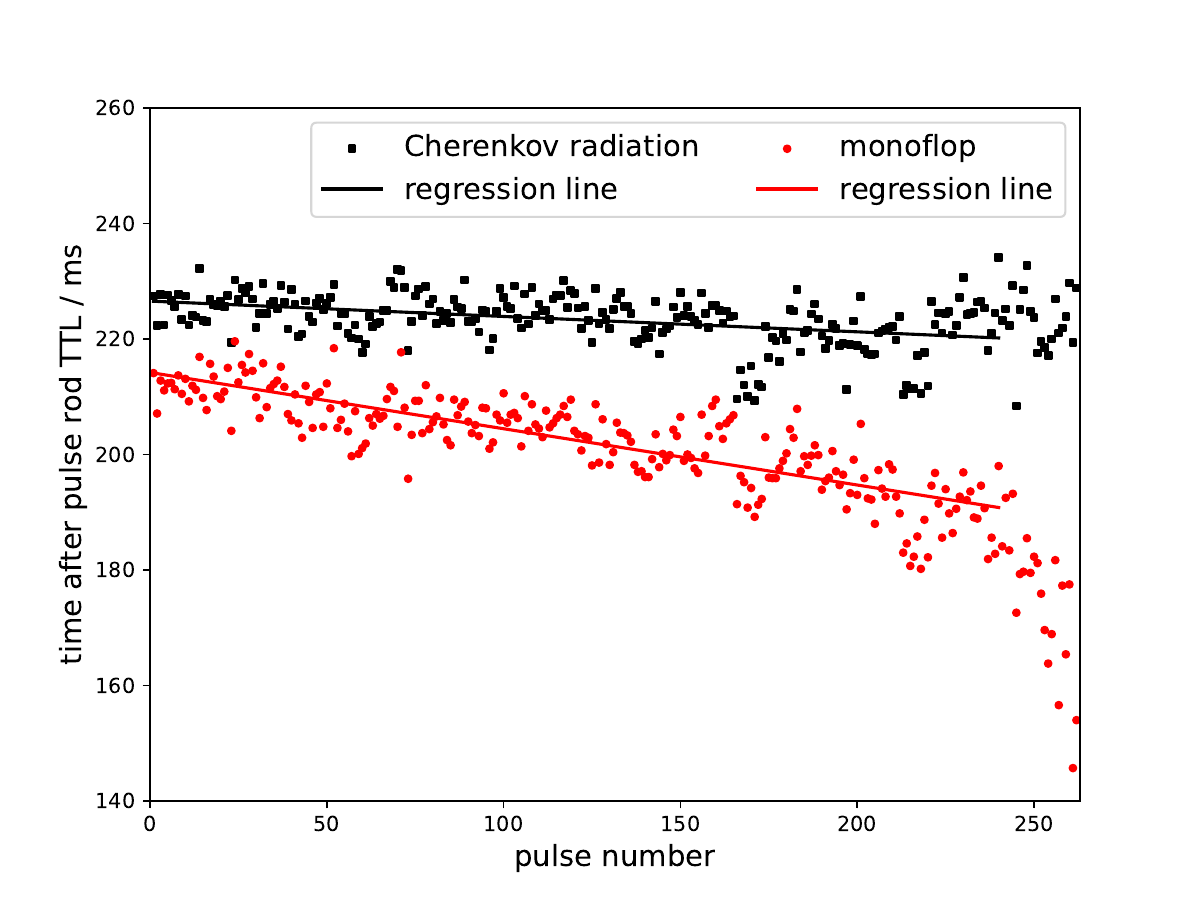}
 	\caption{Time of occurrence of the Cherenkov radiation $t_{\text{Ch}}$ (black) compared to the time of the rising edge of the monoflop output of the Trigger Detector $t_{\text{mono}}$ (red). A linear regression is applied to both data sets. Thresholds were set to $N_{\text{SET}}(U10) = 200$ and $N_{\text{SET}}(U11) = 3000$. Time scale relative to the TTL signal triggered in pulsed mode when the pulse rod hits the stop.}
 	\label{OldSiPM}
 \end{figure}
 \begin{figure}[h]
 	\centering
 	\includegraphics[width=0.75\textwidth]{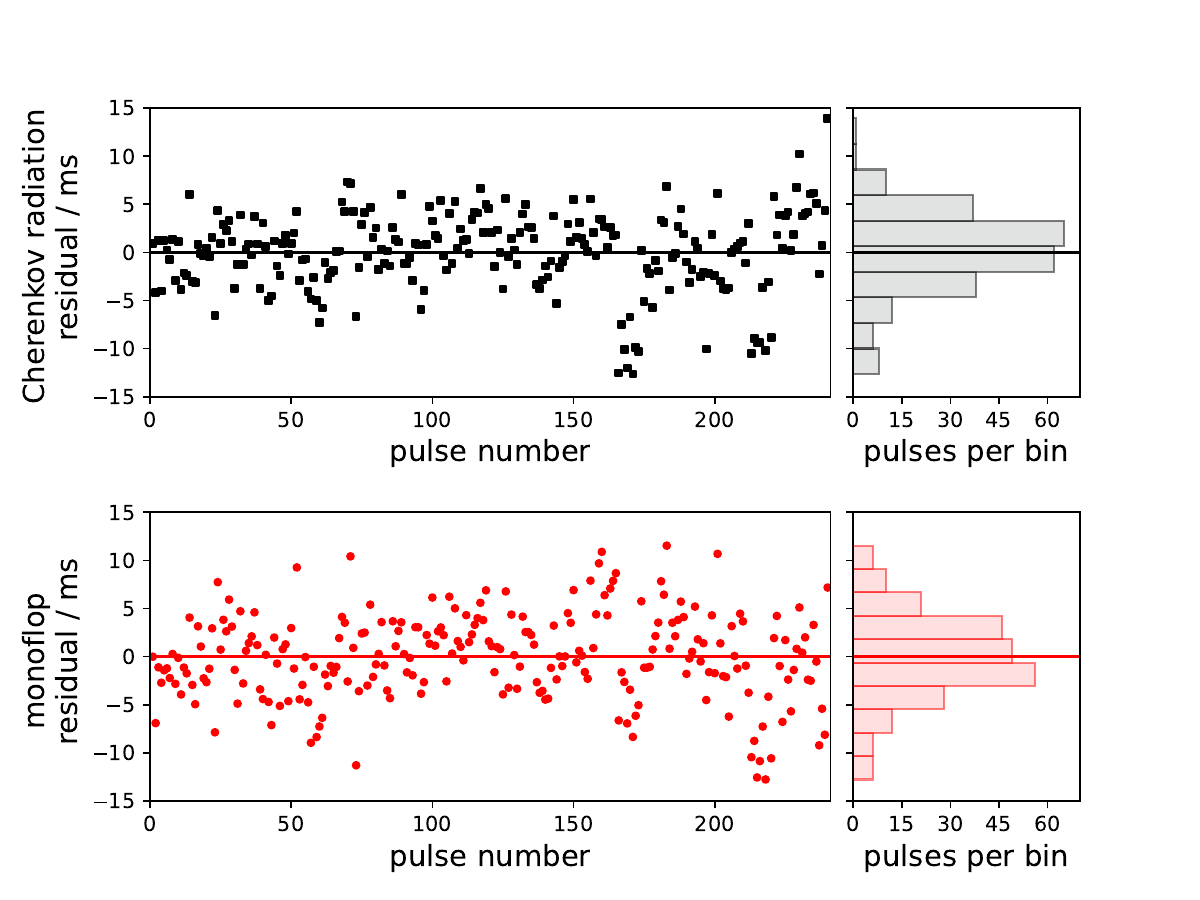}
 	\caption{Residuals of the linear regression for the Cherenkov radiation and the monoflop output of the Trigger Detector. A histogram illustrates the distribution in each case.}
 	\label{LinRegRes}
 \end{figure}
 
 The time of occurrence of the Cherenkov radiation $t_{\text{Ch}}$ is determined by the center of a Gaussian distribution fitted to the data. 
 For comparison, this time and the time of the rising edge of the monoflop output $t_{\text{mono}}$ of the trigger detector are shown in \cref{OldSiPM} for a total of \num{262} reactor pulses with a reactivity of 2\$.
 Both times, $t_{\text{mono}}$ and $t_{\text{Ch}}$, are relative to the TTL signal triggered by the pulse rod.
 The rising edge of the monoflop signal slowly drifts to shorter times until $t_{\text{mono}}$ suddenly drops rapidly around pulse number \num{240}.
 It can be assumed that this drift is due to an increased amplitude of the SiPMs background caused by radiation damage\footnote{This potential damage will be investigated in a future study.} from neutrons or gamma radiation impinging the SiPM during reactor pulses.
 The background amplitude of a new SiPM installed in the Trigger Detector is $\SI{16.3\pm0.7}{\milli\volt}$.
 In contrast the background amplitude of the used SiPM is $\SI{56\pm9}{\milli\volt}$, whereas individual maxima stand out from the background with a peak-to-peak voltage of $\SI{191\pm27}{\milli\volt}$.
 Individual events that were not previously included in the triggering now exceed the input threshold $U_{SET}(U10)$ due to the increased amplitude of the SiPMs background.
 This causes the Trigger Detector to be triggered after shorter times $t_{\text{mono}}$ until finally the background alone is sufficient to trigger the detector permanently.
 In order to reduce the radiation exposure to the SiPM without changing the current setup, the detector can be positioned further away from the reactor, accepting that the range of usable setting parameters will be reduced (see \cref{PositionDependency}).
 
 A linear regression is applied to the data presented in \cref{OldSiPM} up to pulse number \num{240} to compensate for the drift and to enable a comparison of the Trigger Detector data to the Cherenkov radiation.
 The residuals are calculated for both regression lines and shown in \cref{LinRegRes}.
 A histogram illustrates the distribution of the timing in each case. 
 The precision of the times $t_{\text{Ch}}$ and $t_{\text{mono}}$ is given by the standard deviation of the respective distributions as $\upsigma_{\text{Ch}} = \SI{4.1}{\milli\second}$ for the Cherenkov radiation and $\upsigma_{\text{mono}} = \SI{4.5}{\milli\second}$ for the rising edge of the monoflop signal.
 Correlation of the residuals of Cherenkov radiation and monoflop output of the Trigger Detector results in a Pearson correlation coefficient of $r=\num{0.73}$. 
 This correlation of the two data sets can already be seen in \cref{OldSiPM} and \cref{LinRegRes} and is due to the reactor control, since both times $t_{\text{Ch}}$ and $t_{\text{mono}}$ are determined relative to the TTL signal triggered by the pulse rod.
 Thus, the precision described is to be understood as an upper limit in each case.

 \section{Summary and outlook}
 The presented neutron pulse trigger detector is based on the neutron capture reaction in \isotope[10]{B}, and a combination of RC integrator, comparator and monoflop for the generation of a neutron related trigger signal.
 A simulation of the generation of digital trigger signals in combination with measurements with a pulsed LED provides a better understanding of the underlying circuit and confirms the operating principle of the Trigger Detector.
 Suitable parameters for different distances to the research reactor TRIGA Mainz were investigated.
 Optimal parameters for the position closest to the reactor directly at beam port D are determined as $N_{\text{SET}}(U10) = 200$ and $N_{\text{SET}}(U11) = 3000$.
 An upper limit for the precision of the timing provided by the Trigger Detector was determined relative to the reactor control systems as $\upsigma_{\text{mono}} = \SI{4.5}{\milli\second}$.
 Since timing jitters in the reactor control can be resolved by the Trigger Detector it is indicated that the presented detector design is suitable for precise, facility-independent experimental timing coinciding with pulsed neutron production at the research reactor TRIGA Mainz and possibly at other pulsed sources.
 With this relative temporal precision achieved, which can be understood as an upper limit, the Trigger Detector could possibly also be used at accelerator-driven pulsed neutron sources, such as the SNS or the ESS.
 
 An upgrade of the Trigger Detector is under development in which the multilayer structure is divided into two spatially separated components of \isotope[10]{B}-coated scintillation foil and SiPM, coupled by a light guide with suitable geometry.
 With this arrangement, the detector electronics, especially the SiPM, can be removed from the beam and additionally shielded from neutrons and gamma radiation, while the \isotope[10]{B}-coated scintillator foil can remain directly on the beam line in front of beam port D of the reactor.

\section*{Data availability} 
 All relevant data as well as the circuit diagram and production files of the Trigger Detector board are available from the corresponding authors upon request.

 \section*{Acknowledgements}
 The authors acknowledge the excellent support of the TRIGA Mainz electronics workshop personnel and the reactor operation crews.
 Support by the Cluster of Excellence “Precision Physics, Fundamental Interactions, and Structure of Matter” (PRISMA+ \& EXC 2118/1) funded by the German Research Foundation (DFG) within the German Excellence Strategy (Project ID 39083149) is acknowledged.
  
 \bibliographystyle{unsrt}
 \bibliography{mybib}
\end{document}